\documentclass[conference]{IEEEtran}
\IEEEoverridecommandlockouts

\usepackage{cite}
\usepackage{amsmath,amssymb,amsfonts}
\usepackage{graphicx}
\usepackage{textcomp}
\usepackage{xcolor}

\usepackage{amsthm}
\usepackage[ruled,vlined,linesnumbered]{algorithm2e}
\usepackage{url}
\usepackage{graphicx}
\usepackage{comment}
\usepackage{booktabs}
\usepackage{caption}
\usepackage{balance}
\usepackage{float}

\usepackage{threeparttable}
\usepackage{array} 
\usepackage{graphicx}
\usepackage{subcaption}
\usepackage{algpseudocode}
\usepackage{multirow}
\usepackage{color}

\def\BibTeX{{\rm B\kern-.05em{\sc i\kern-.025em b}\kern-.08em
    T\kern-.1667em\lower.7ex\hbox{E}\kern-.125emX}}
\begin{document}

\title{DRIFT: Direct Reduced Fourier Transforms for Distributed Spectral Neural Operators}

\author{

Sana Taghipour Anvari,
David Kaeli
}

\maketitle

\begin{abstract}
Fourier Neural Operators (FNOs) learn solution operators for partial differential equations and offer orders of magnitude speedup over traditional numerical solvers at inference time, which makes them attractive surrogates for high-resolution computational physics. Scaling FNOs to high-resolution spatial grids requires distributing the data across GPUs, but the distributed FFT at the core of each spectral layer requires multiple dense all-to-all collectives that communicate the full spatial tensor, only for most coefficients to be discarded immediately. We introduce the Distributed Truncated Spectral Transform (DTST), which reverses this order. Each GPU computes only a small subset of frequency modes used by the spectral convolution locally via a partial DFT, and two collectives combine the results with a payload that depends only on this mode count, not the spatial resolution. DTST produces spectral coefficients identical to the standard distributed FFT with truncation, while providing both spatial data parallelism and spectral weight model parallelism. We present DRIFT, a GPU implementation of DTST for distributed Fourier Neural Operators, using separable per-dimension basis matrices and efficient GPU-to-GPU communication. On a 3D+time FNO across 4--32 GPUs, on up to 8 nodes (4 GPUs/node), DRIFT achieves a forward-pass speedup of 38--64$\times$ and a 37$\times$ training speedup over the distributed FNO baseline, reducing communication time from 97\% to under 6\% of the forward-pass time, with growing speedups at higher resolution.
\end{abstract}


\section{Introduction}
Fourier Neural Operators (FNOs)~\cite{FNO} have emerged as a leading architecture for learning solution operators of partial differential equations (PDEs). FNOs learn mappings between function spaces that generalize across discretizations at low computational cost by parameterizing the kernel integral operator in Fourier space and using the Fast Fourier Transform (FFT). A defining feature of FNOs is their associated mode truncation, which means only a small number of frequency modes are retained per spatial dimension, and the majority of the spectral coefficients are discarded. In practice, the number of retained modes is orders of magnitude smaller than the spatial resolution.

Scaling FNOs to high-resolution, high-dimensional problems in scientific computing applications requires distributing both data and model parameters across multiple GPUs, as GPUs are widely used accelerators for scientific ML workloads due to their high throughput for FFTs and dense linear algebra operations that dominate FNO layers. Channel-parallel approaches, such as the Adaptive Fourier Neural Operator~\cite{AFNO}, avoid spatial communication by replicating the full grid on every GPU. But this strategy cannot be applied when the spatial grid exceeds single-device memory, as is the case for high-resolution 3D and 4D problems~\cite{FourCastNet}. For such problems, spatial domain decomposition is the only viable path. Using a spatially distributed FNO~\cite{GradyDFNO} enables predictions over billions of variables on a large number of GPUs. Still, the distributed FFTs, which are at the core of each spectral layer, inherently require multiple all-to-all collectives to redistribute data along each transform dimension globally. These communication operations become the dominant bottleneck and consume most of the forward-pass time at scale, while the communication pattern they impose is fundamentally wasteful. The distributed FFT computes all frequency coefficients, redistributes the full spatial tensor across GPUs, and subsequently truncates and discards almost everything that was computed. The fraction of communicated data that actually contributes to the output is $(k_{\max}/N)^d$, which becomes a rapidly decreasing percentage of processing as we increase the resolution $N$ and the dimensionality $d$ of the problem.

The key insight of this work is that mode truncation does not need to follow the distributed data exchange, and that the order of these two operations can be reversed. Rather than computing the full distributed FFT and discarding most of the results, we compute only the needed frequency coefficients locally via a partial Discrete Fourier Transform (DFT) and combine these partial contributions across GPUs through an AllReduce on the compact spectral tensor. This reordering transforms the communication pattern of each spectral layer from multiple dense all-to-all collectives over the full spatial tensor to two collectives whose payload depends only on the number of retained modes, not on the spatial resolution or GPU count. We formalize this idea as the Distributed Truncated Spectral Transform (DTST), a general communication primitive for computing a subset of the DFT coefficients from spatially distributed data. DTST is exact and symmetric, such that forward and backward passes incur identical communication costs. We instantiate DTST within the FNO architecture as DRIFT (Direct Reduced Fourier Transforms), a distributed reduced-communication inverse and forward transform. DRIFT exploits the separability of the multi-dimensional DFT through sequential per-dimension matrix multiplies, and applies the partial DFT independently along each distributed dimension. After the AllReduce, each GPU selects its partition of the retained spectrum and applies only its local shard of the spectral weights, which unifies data parallelism over the spatial domain, with model parallelism over the spectral parameters. An AllGather reassembles the convolved spectrum for the inverse transform.

The main contributions of our approach include:
\begin{enumerate}
    \item We introduce the Distributed Truncated Spectral Transform (DTST), a communication primitive that replaces the distributed FFT with local partial DFTs and an AllReduce/AllGather pair on the retained spectral coefficients. DTST is exact, backend-agnostic, and applicable to any spectral method that uses mode truncation.
    \item We show that DTST reduces the per-layer communication volume from $O(N^d / P)$ to $O(k_{\max}^d)$, independent of spatial resolution, with $O(\log P)$ latency scaling. The reduction factor grows exponentially with problem dimensionality.
    \item We develop an efficient GPU implementation using separable per-dimension partial DFT via NVIDIA cuBLAS and GPU-aware MPI, and show that progressive compression across dimensions makes the partial DFT more efficient than full cuFFT transforms, despite replacing an $O(N \log N)$ algorithm with $O(KN)$ matrix multiplications.
    \item We evaluate DRIFT on distributed FNO inference and training across up to 32 GPUs, demonstrating a 38--64$\times$ forward-pass speedup and a 37$\times$ training speedup over the baseline distributed  FNO, with comparable convergence and spectral coefficients verified to be exact.
\end{enumerate}

\section{Related Work}
{\bf Neural Operators:}
Neural operators learn mappings between function spaces and allow fast surrogates for parametric PDEs that generalize across spatial resolutions~\cite{NO}. DeepONet~\cite{DeepONet} encodes input functions and query locations through separate branch and trunk networks, with physics-informed variants incorporating PDE constraints directly into training~\cite{PDeepONet}. The Fourier Neural Operator (FNO)~\cite{FNO} parameterizes the integral of the kernel in Fourier space, using FFTs for efficient global convolution, and with mode truncation for regularization and computational efficiency.  FNO has been extended through factorized spectral representations~\cite{FFNO}, multi-grid tensorization~\cite{MGTFNO}, Tucker decomposition of spectral weights~\cite{TuckerFNO}, and decomposition of the latent representation into separable 1D transforms~\cite{LiYeDFNO}. The Adaptive Fourier Neural Operator (AFNO)~\cite{AFNO} replaces the dense spectral weight tensor with a block-diagonal channel mixer for efficient token mixing within vision transformers.

FourCastNet~\cite{FourCastNet} deployed AFNO for global weather prediction on up to 3,808 GPUs using channel parallelism, where certain operations require communication across model-parallel ranks. TurboFNO~\cite{TurboFNO} fuses the FFT-GEMM-iFFT pipeline into a single GPU kernel and achieves significant speedups. Grady et al.~\cite{GradyDFNO} introduced the first spatially distributed FNO using a linear-algebraic framework for model parallelism~\cite{DistDL}, enabling predictions over billions of variables on up to 512 GPUs. Their approach distributes the spatial grid across GPUs and relies on the distributed FFT for spectral transforms. At scale, the authors report that the all-to-all redistributions required by the distributed FFT consume the vast majority of the forward-pass time, limiting scaling efficiency.

{\bf Distributed FFT Scalability:}
The communication bottleneck in distributed FFTs is well studied. Czechowski et al.~\cite{Czechowski} analyzed the communication complexity of 3D FFTs and showed that all-to-all transposes dominate runtime at large scales. Ayala et al.~\cite{AyalaScale} demonstrated that all-to-all communication fails to scale on large systems such as Fugaku, while the heFFTe library~\cite{HeFFTe} reported that MPI communication consumes over 97\% of GPU-accelerated FFT runtime on Summit~\cite{AyalaGPU}. Distributed FFT libraries, including PFFT~\cite{PFFT}, 2DECOMP\&FFT~\cite{2DECOMP}, and heFFTe~\cite{HeFFTe}, have optimized pencil and slab decompositions, but all fundamentally require global data redistribution through all-to-all collectives. 

Communication avoidance approaches have been explored for related transforms, such as FMM-accelerated FFTs~\cite{FMMFFT} that reduce communication by exploiting the hierarchical structure of multipole expansions. DaggerFFT~\cite{DaggerFFT} introduced a task-based distributed FFT framework that pipelines communication with computation through dynamic scheduling and achieves overlap between redistributions and local FFT stages, but the underlying communication pattern still requires global data redistribution. Popovici et al.~\cite{Popovici} used SMT-based optimization to automatically generate efficient mappings for distributed multi-dimensional Fourier operations, minimizing communication steps across decomposition stages.

{\bf Communication Models:}
Our analysis is based on established models for MPI collectives. The $\alpha$-$\beta$ cost model~\cite{Kumar} characterizes collective operations in terms of latency $\alpha$ and inverse bandwidth $\beta$. Optimal allreduce algorithms achieve $O(\alpha \log P + \beta M)$ cost~\cite{Chan, Pjesivac}, compared to all-to-all cost that scales linearly in $P$~\cite{Hoefler}. DRIFT's communication advantage follows directly from this asymptotic difference.

Despite progress in both neural operator architectures and distributed FFT implementations, a fundamental challenge remains unaddressed. When high-resolution 3D and 4D spatial grids exceed the memory present on a single GPU, the spectral layers that define FNO become communication-bound. Approaches that replicate the full spatial grid on every GPU cannot scale to such problem sizes, while spatially distributed approaches inherit the scaling limitations of the distributed FFT.

\section{Background and Motivation}
Fourier Neural Operators (FNO)~\cite{FNO} learn mappings between function spaces by parameterizing an integral operator in Fourier space. The input function $a(x)$ is first lifted to a higher-dimensional channel representation $v^{(0)}(x) \in \mathbb{R}^{d_v}$ via a pointwise network $P$, passed through $L$ Fourier layers, and projected back to the output dimension via a pointwise network $Q$. Given an input field $v^{(\ell)}$, a Fourier layer applies:
\begin{equation}
v^{(\ell+1)}(x) = \sigma\Big(W^{(\ell)} v^{(\ell)}(x)+\mathcal{F}^{-1}\big(R_\theta^{(\ell)} \cdot \mathcal{F}(v^{(\ell)})\big)(x)\Big)
\label{eq:fno_layer}
\end{equation}
where $W^{(\ell)}$ is a point-wise linear map and $R_\theta^{(\ell)}$ is a complex-valued, learnable filter that acts on Fourier coefficients. A defining feature of FNOs is their associated mode truncation, where only the lowest $k_{\max}$ frequencies per spatial dimension are retained. Let $S_j = \{0, \dots, k_{\max}-1\} \cup \{N_j - k_{\max}, \dots, N_j - 1\}$ denote the retained indices along dimension $j$. These correspond to the $k_{\max}$ lowest positive and negative frequency modes, so $|S_j| = 2k_{\max}$. The spectral update is restricted to $\mathbf{k} \in S_1 \times \cdots \times S_d$, while all other coefficients are discarded. The number of modes used, therefore, scales as $(2k_{\max})^d$, independent of the spatial resolution $N$. In practical settings, $k_{\max} \ll N$, often by one to two orders of magnitude~\cite{FNO,FNOEX}.

\begin{figure}[t]
\centering
\includegraphics[width=\columnwidth]{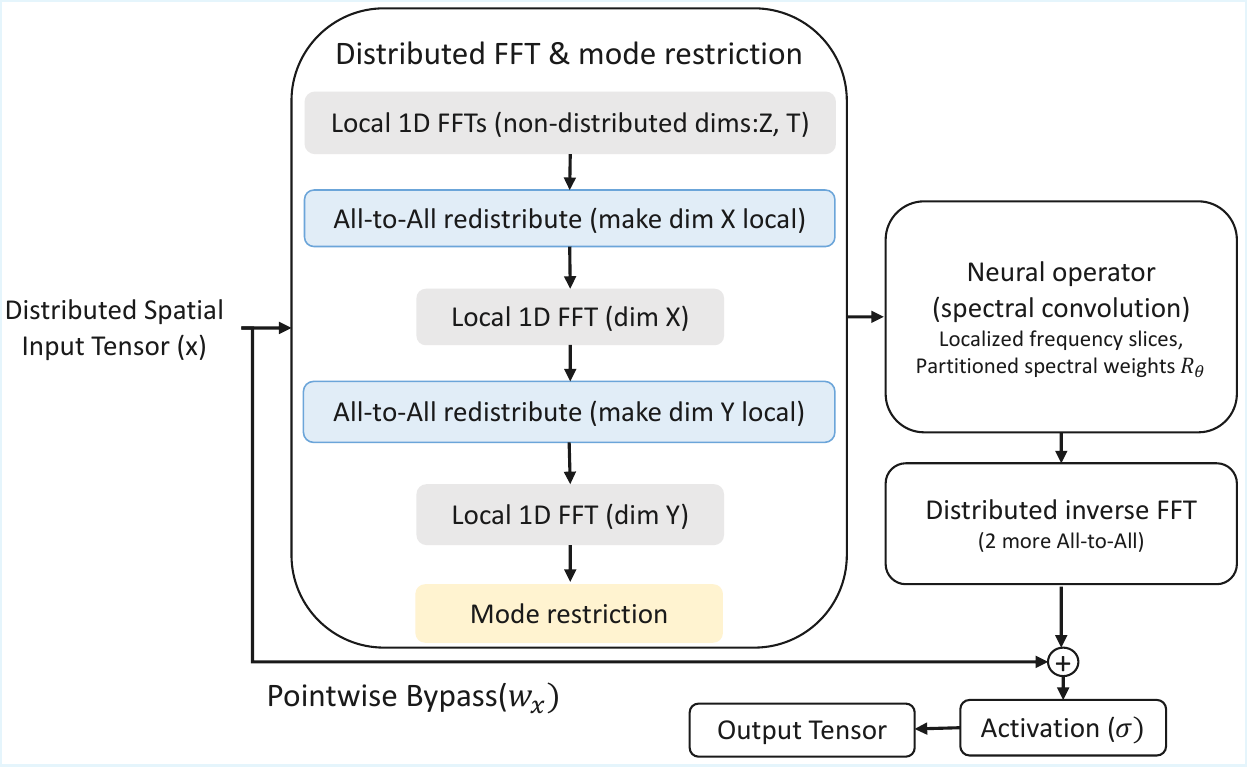}
\caption{Standard distributed FNO layer.}
\label{fig:dfno_arch}
\end{figure}

At high resolutions (e.g., spatial grids of $128^3$ or larger with multiple fields and timesteps), the spatial grid exceeds the memory capacity of a single GPU and must be distributed across multiple GPUs. Grady et al.~\cite{GradyDFNO} introduced the first spatially distributed FNO using domain decomposition with repartition operators. Each GPU owns a contiguous block of the spatial domain and repartition collectives redistribute the tensor so that each FFT dimension is locally present before transformation. Each FNO block (see Figure~\ref{fig:dfno_arch}) applies the iterative update in Eq.~\eqref{eq:fno_layer}, where the distributed spectral convolution replaces the standard FFT with a sequence of repartition operators and local transforms:
\begin{equation}
\mathcal{F}_{\text{dist}}\, v = 
\mathcal{F}_{\mathcal{I}_k}\, 
T_{\{P_{\mathcal{I}_{k-1}}\} \to \{P_{\mathcal{I}_k}\}} 
\cdots\, 
\mathcal{F}_{\mathcal{I}_1}\, 
T_{\{P\} \to \{P_{\mathcal{I}_1}\}}\, v
\label{eq:dfno_dfft}
\end{equation}
where each $T_{\{P\} \to \{Q\}}$ is a repartition operator (all-to-all collective) that redistributes the tensor from partition $P$ to partition $Q$, and $\mathcal{F}_{\mathcal{I}_j}$ applies a local FFT along the dimensions in index set $\mathcal{I}_j$ that are now fully local on each GPU. The distributed spectral convolution can be computed as:
\begin{equation}
(\mathcal{S}_{\text{dist}}\, v^{(\ell)})(x) = 
\mathcal{F}^{\top}_{\text{dist}}
\big(R_\theta^{(\ell)} \cdot 
(\mathcal{F}_{\text{dist}}\, v^{(\ell)})\big)(x)
\label{eq:dfno_spec}
\end{equation}
which is added to the point-wise linear bypass $W^{(\ell)} v^{(\ell)}(x)$ and passed through the activation $\sigma$. The spatial tensor is redistributed via repartition collectives so that each GPU holds the full extent of a group of spatial dimensions, and local FFTs are applied along those dimensions. A second repartition makes the remaining dimensions local for the next group of FFTs. The inverse transform reverses the process with another two repartitions. Because each repartition leaves a different subset of frequency modes on each GPU, the spectral weights $R_\theta$ must be partitioned to match, coupling model parallelism to the communication pattern. In total, each FNO block requires 4 all-to-all repartitions, each moving $O(N^d \cdot d_v / P)$ data per GPU, where $d_v$ is the channel width.

\begin{figure}[t]
\centering
\includegraphics[width=\columnwidth]{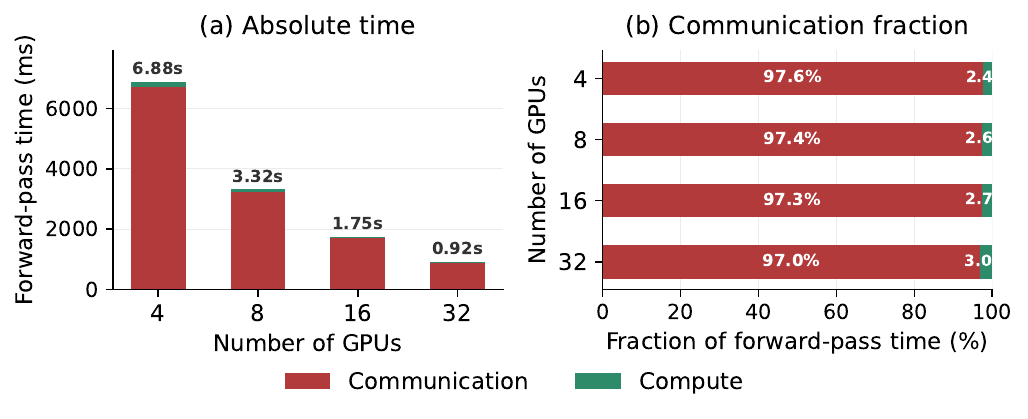}
\caption{Communication vs. compute breakdown of the baseline DFNO forward pass on a $128{\times}128{\times}64$ grid with modes $(8,8,8,16)$, $d_v{=}20$, 4 blocks, across 4--32 GPUs, 4 GPUs per node, with one rank per GPU.}
\label{fig:dfno_profile}
\end{figure}
\begin{figure*}[t]
\centering
\includegraphics[width=0.9\textwidth]{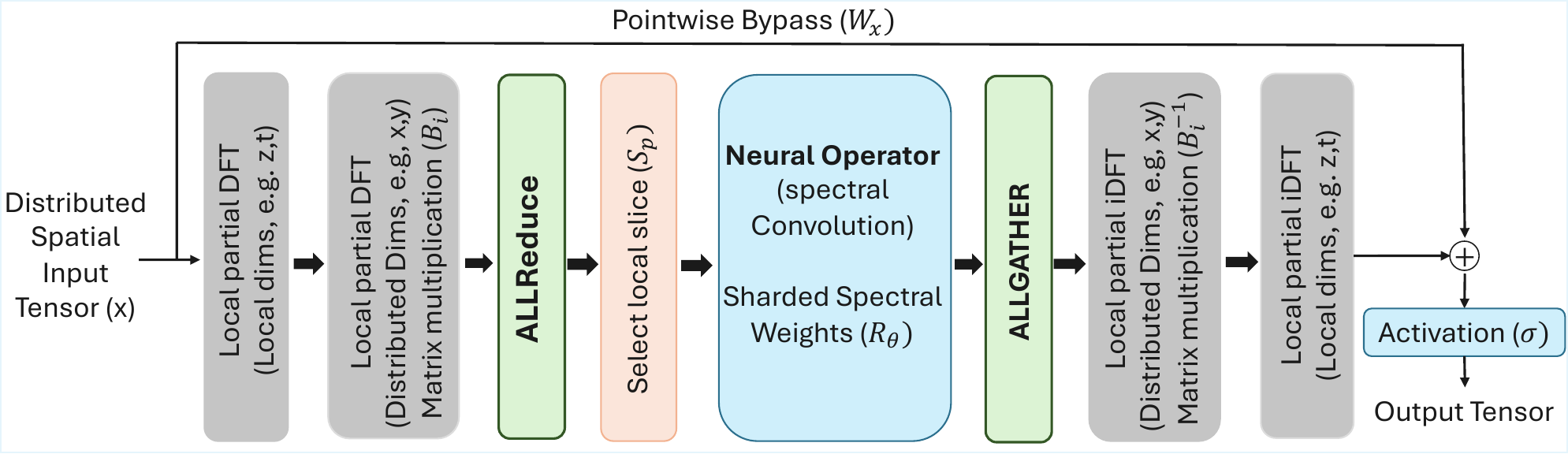}
\caption{The DRIFT layer architecture. Each GPU applies a local partial DFT first along the fully local dimensions (Z, T), then along the partitioned dimensions (X, Y) via a basis-matrix multiplication $B_i$. An AllReduce sums partial spectra across all GPUs. Each GPU selects its partition of the retained spectrum and multiplies it by its local shard of the spectral weights $R_\theta$. An AllGather reassembles the full convolved spectrum. The inverse partial DFT reconstructs each GPU's spatial output via $B_i^{-1}$ without additional communication. A point-wise bypass $W_x$ and activation $\sigma$ complete the block.}
\label{fig:architecture}
\end{figure*}

As GPU compute throughput continues to outpace interconnect bandwidth, these collectives increasingly dominate runtime. Figure~\ref{fig:dfno_profile} profiles a baseline distributed FNO using a grid size of $128 \times 128 \times 64$, modes $(8,8,8,16)$, width 20, and 4 blocks. Communication accounts for over 97\% of the forward-pass time when run on 4--32 GPUs across 1--8 nodes (4 GPUs per node), consistent with prior reports on distributed FFT at scale~\cite{AyalaGPU, AyalaScale}. The compute fraction never exceeds 3.3\%, which confirms that the spectral weight multiplication is negligible relative to the all-to-all collectives. Yet the distributed FFT computes and communicates all $N^d$ coefficients, but only retains $(2k_{\max})^d$ of them. The useful fraction of communicated data is:
\begin{equation}
\frac{V_{\text{used}}}{V_{\text{communicated}}}
=
\frac{(2k_{\max})^d}{N^d}
=
\left(\frac{2k_{\max}}{N}\right)^d
\label{eq:waste_ratio}
\end{equation}
which rapidly approaches zero as the resolution and dimension of the computation increase. Thus, communication complexity is governed by $\Theta(N^d)$, while the useful spectral information scales with $\Theta(k_{\max}^d)$.

\section{The Distributed Truncated Spectral Transform (DTST)}
We introduce the \emph{Distributed Truncated Spectral Transform} (DTST), a communication primitive for computing a subset of the discrete Fourier coefficients from spatially distributed data without a distributed FFT. DTST replaces the distributed FFT in spectral methods that retain and operate only a small fraction of the frequency modes.

\subsection{Problem Setting}
Consider a $d$-dimensional tensor $x \in \mathbb{C}^{N_1 \times \cdots \times N_d}$ distributed across $P$ GPUs via spatial domain decomposition. GPU~$p$ holds a contiguous block $x_p$ with local extents $N_i^{\mathrm{loc}} = N_i / P_i$, where $P = \prod_i P_i$. Let $\mathcal{S} = \mathcal{S}_1 \times \cdots \times \mathcal{S}_d$ denote the set of $M = \prod_i |\mathcal{S}_i|$ frequency modes to be computed, with $M \ll \prod_i N_i$. The standard approach computes the full distributed FFT over all $\prod_i N_i$ coefficients via multiple dense all-to-all collectives, then discards all but the $M$ desired modes. DTST reverses this workflow by computing only the desired modes locally and aggregates them across GPUs through a global summation.

\subsection{Mathematical Definition}
The standard DFT of $x$ at frequency $\mathbf{k}$ is a summation over the full spatial domain $\Omega = \{0,\ldots,N_1{-}1\} \times \cdots \times \{0,\ldots,N_d{-}1\}$:
\begin{equation}
\hat{X}[\mathbf{k}] = \sum_{\mathbf{j} \in \Omega} x[\mathbf{j}] \prod_{i=1}^{d} W_{N_i}^{k_i j_i}
\label{eq:full_dft}
\end{equation}
where $W_N = e^{-2\pi i / N}$. Since the GPUs partition $\Omega$ into disjoint subsets, $\Omega = \bigsqcup_{p=0}^{P-1} \mathrm{dom}(p)$, the global summation can be split into independent per-GPU sums, one over each local subdomain:
\begin{equation}
\hat{X}[\mathbf{k}] = \sum_{p=0}^{P-1} \underbrace{ \sum_{\mathbf{j} \in \mathrm{dom}(p)} x[\mathbf{j}] \prod_{i=1}^{d} W_{N_i}^{k_i j_i} }_{\hat{X}_p[\mathbf{k}]}, \quad \forall\, \mathbf{k} \in \mathcal{S}
\label{eq:dtst_decompose}
\end{equation}
Each GPU independently evaluates its local partial spectrum $\hat{X}_p[\mathbf{k}]$ for only the retained modes $\mathbf{k} \in \mathcal{S}$, and the global coefficients are obtained by summing the partial contributions across all GPUs:
\begin{equation}
\hat{X}[\mathbf{k}] = \sum_{p=0}^{P-1} \hat{X}_p[\mathbf{k}]
\label{eq:dtst_reduce}
\end{equation}
This produces spectral coefficients identical to the full distributed FFT, followed by a truncation step. Because the DFT is linear, the order of truncation and aggregation can be reversed without affecting the result.

\subsection{Communication Cost}
We analyze communication costs using the $\alpha$--$\beta$ model~\cite{hockney1994communication, Chan}. This model characterizes the time to send a message of size $m$ bytes between two nodes as $T = \alpha + \beta \, m$, where $\alpha$ is the per-message latency and $\beta$ is the per-byte transfer cost (seconds per byte). Under this model, the cost of aggregating partial spectra via an AllReduce is:
\begin{equation}
T_{\mathrm{DTST}} = 2\lceil\log_2 P\rceil \cdot \alpha + 2\beta\,\frac{M \cdot c \cdot (P{-}1)}{P},
\label{eq:cost_dtst}
\end{equation}
where $c$ is the payload per-mode in bytes. The factor of~$2$ in the latency term reflects the two phases of an AllReduce (reduce-scatter + allgather), and the bandwidth term accounts for the $2M c (P{-}1)/P$ bytes transferred during AllReduce. By contrast, the distributed FFT communicates the full spatial tensor via all-to-all repartitions at a cost of:
\begin{equation}
T_{\mathrm{FFT}} = \alpha\,(P{-}1) + \beta\,\frac{\prod_i N_i}{P}\,c
\label{eq:cost_fft}
\end{equation}
per repartition~\cite{Hoefler}. The ratio of these costs gives a per-repartition communication reduction factor of:
\begin{equation}
\frac{T_{\mathrm{FFT}}}{T_{\mathrm{DTST}}} \;\propto\; \frac{(P{-}1)}{2\lceil\log_2 P\rceil} \cdot \frac{\prod_i N_i}{P \cdot M}
\label{eq:reduction_factor}
\end{equation}
which grows with both spatial resolution and GPU count. The key insight is that DTST communicates a payload proportional to $M$ with $O(\log P)$ latency, while the distributed FFT communicates a payload proportional to $\prod_i N_i / P$, with $O(P)$ latency. For spectral methods where $M \ll \prod_i N_i$, this represents a fundamental reduction in communication volume and scaling cost.

\section{DRIFT: DTST for Distributed Fourier Neural Operators}
We apply DTST to the distributed Fourier Neural Operator (DFNO) to build DRIFT, a distributed FNO layer that replaces the four all-to-all repartitions per block in a DFNO forward pass with two collectives per block, whose payload depends only on the mode count. Figure~\ref{fig:architecture} illustrates the DRIFT architecture.

\begin{figure*}[t]
\centering
\includegraphics[width=\textwidth]{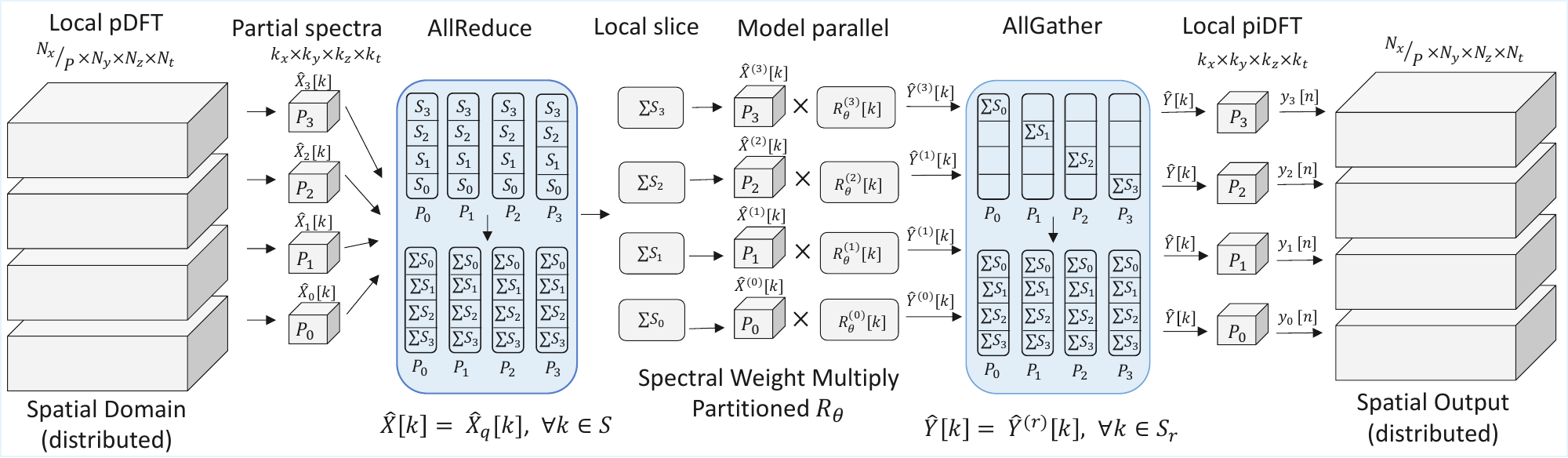}
\caption{Data-centric view of the DRIFT pipeline with $P=4$ GPUs. Each GPU computes a local partial spectrum from its spatial partition. Each AllReduce sums the partial contributions so that every GPU holds the identical full spectrum. Each GPU selects its partition $\mathcal{S}_p$ and applies its local shard of $R_\theta$. The AllGather reassembles the fully convolved spectrum. The inverse partial DFT reconstructs each GPU's spatial output without additional communication.}
\label{fig:data_flow}
\end{figure*}

\subsection{Distributed Spectral Convolution with Partitioned Weights}
DRIFT uses an AllReduce to sum partial spectra across all GPUs and produces the full spectrum of the $M$ modes on every GPU. Each GPU then selects its disjoint partition of the spectrum for a partitioned spectral convolution (see Fig.~\ref{fig:data_flow}).

Let $\mathcal{S} = \bigsqcup_{p=0}^{P-1} \mathcal{S}_p$ be an equal partition of the $M$ modes into $P$ disjoint subsets, each of size $M/P$. The AllReduce sums partial contributions so that every GPU holds the full spectrum:
\begin{equation}
\hat{X}[\mathbf{k}]
= \sum_{q=0}^{P-1} \hat{X}_q[\mathbf{k}],
\quad \forall\, \mathbf{k} \in \mathcal{S}
\label{eq:allreduce}
\end{equation}
Each GPU~$p$ selects its local partition $\mathcal{S}_p$ and applies its local shard of the learnable spectral convolution weights $R_\theta^{(p)} \in \mathbb{C}^{|\mathcal{S}_p| \times d_v \times d_v}$:

\begin{equation}
\hat{Y}^{(p)}[\mathbf{k}] = R_\theta^{(p)}[\mathbf{k}] \, \hat{X}[\mathbf{k}], \quad \mathbf{k} \in \mathcal{S}_p
\label{eq:partitioned_conv}
\end{equation}
An AllGather reassembles the full convolved spectrum on every GPU:
\begin{equation}
\hat{Y}[\mathbf{k}]
= \hat{Y}^{(r)}[\mathbf{k}],
\quad \mathbf{k} \in \mathcal{S}_r,
\quad r = 0, \ldots, P{-}1
\label{eq:allgather}
\end{equation}
The local inverse partial DFT then reconstructs only the spatial points owned by GPU~$p$:
\begin{equation}
y_p[\mathbf{n}]
= \frac{1}{\prod_i N_i}
\sum_{\mathbf{k} \in \mathcal{S}}
\hat{Y}[\mathbf{k}]
\prod_{i=1}^{d} W_{N_i}^{-k_i n_i},
\quad \mathbf{n} \in \mathrm{dom}(p)
\label{eq:dtst_inverse}
\end{equation}
Because $\hat{Y}[\mathbf{k}]$ is available on every GPU after the AllGather, this inverse requires no communication.

{\bf Weight partitioning and gradient communication:}
After the AllReduce, each GPU selects its partition $\mathcal{S}_p$ of $M/P$ modes and stores only its local shard of the spectral weights $R_\theta^{(p)} \in \mathbb{C}^{M/P \times d_v \times d_v}$. AllReduce transfers $2Mc(P{-}1)/P$ bytes and AllGather transfers $Mc(P{-}1)/P$ bytes, where $c = d_v \cdot \mathrm{sizeof(complex)}$. This yields a total communication volume of:
\begin{equation}
V_{\mathrm{DRIFT}} = 3\,M \cdot c \cdot \frac{P{-}1}{P} \cdot L
\label{eq:drift_volume}
\end{equation}
per forward pass across $L$ blocks. This volume is independent of the spatial resolution $N_i$ and the GPU count~$P$ (for $P \gg 1$). Because each GPU computes weight gradients $\partial \mathcal{L} / \partial R_\theta^{(p)}$ locally on its own partition, no additional gradient synchronization is needed, unlike standard data-parallel training which requires an AllReduce over weight gradients. When $P = 1$, no communication is performed and DRIFT reduces to the standard single-GPU FNO without overhead.

{\bf Communication model for DRIFT:}
Applying the DTST communication cost (Eq.~\eqref{eq:cost_dtst}) plus the AllGather cost to the FNO forward pass with L blocks:
\begin{equation}
T_{\mathrm{DRIFT}} = L\left[3\lceil\log_2 P\rceil \cdot \alpha + 3\beta\,\frac{M \cdot c \cdot (P{-}1)}{P}\right]
\label{eq:cost_drift}
\end{equation}

The distributed FFT baseline performs $4L$ all-to-all repartitions (4 per block), giving:
\begin{equation}
T_{\mathrm{DFFT}} = 4L\left[\alpha\,(P{-}1) 
+ \beta\,\frac{\prod_i N_i}{P}\,c\right]
\label{eq:cost_dfft}
\end{equation}
Table~\ref{tab:ab_validation} validates both communication models against measured collective times. The all-to-all model achieves ${<}3.3\%$ prediction error across all GPU counts. For the AllReduce and AllGather, per-block fits on the inter-node configurations ($P \geq 8$, 2--8 nodes) give $T_{\mathrm{AR}} = 1.07 + 0.29\,\log_2 P$ and $T_{\mathrm{AG}} = 0.44 + 0.54\,\log_2 P$ (ms/block), which confirms the $\log_2 P$ scaling predicted by the theoretical model with prediction errors below 8\%.

\begin{table}[t]
\centering
\caption{Communication model validation. Measured collective time (ms) per forward pass versus $\alpha$--$\beta$ model predictions on a $128{\times}128{\times}64$ grid with modes $(8,8,8,16)$, $d_v{=}20$, $L{=}4$ blocks, 4 GPUs per node, one rank per GPU.}
\label{tab:ab_validation}
\small
\begin{tabular}{r rr r rr r}
\toprule
& \multicolumn{3}{c}{\textbf{All-to-all (repartitions)}}
& \multicolumn{3}{c}{\textbf{AR + AG}} \\
\cmidrule(lr){2-4} \cmidrule(lr){5-7}
\textbf{P}
& \textbf{Pred.} & \textbf{Meas.} & \textbf{Err.}
& \textbf{Pred.} & \textbf{Meas.} & \textbf{Err.} \\
\midrule
8  & 3343 & 3236 & 3.3\%   &  16.0 &  15.0 & 6.7\% \\
16  & 1689 & 1705 & 1.0\%   &  19.3 &  20.9 & 7.7\% \\
32  &  879 &  891 & 1.3\%   &  22.6 &  21.6 & 4.6\% \\
\bottomrule
\end{tabular}
\end{table}

\subsection{Computing the Local Partial Spectrum}
Each GPU evaluates Eq.~\eqref{eq:dtst_decompose} via matrix multiplications with precomputed DFT basis matrices. For each spatial dimension~$i$, the basis matrix $B_i \in \mathbb{C}^{|\mathcal{S}_i| \times N_i^{\mathrm{loc}}}$ has entries
\begin{equation}
(B_i)_{k,j} = W_{N_i}^{k(p_i N_i^{\mathrm{loc}} + j)},
\quad k \in \mathcal{S}_i
\label{eq:basis_matrix}
\end{equation}
where $p_i$ is the GPU coordinate along dimension~$i$ and the phase offset $p_i N_i^{\mathrm{loc}}$ encodes the global position of each local element. For non-distributed dimensions, where each GPU holds the full extent, the offset is zero and the basis matrix reduces to a standard DFT submatrix, selecting the retained modes.

The partial DFT is applied as sequential matrix multiplications along each dimension (see Fig.~\ref{fig:data_flow}). Each spatial axis is mapped from its full local extent~$N_i^{\mathrm{loc}}$ to $|\mathcal{S}_i|$ modes. This progressively shrinks the tensor. Non-distributed dimensions are processed first, as they require no phase offsets and immediately compress the tensor. Distributed dimensions are processed last, producing the local partial spectrum~$\hat{X}_p$ that is summed across GPUs via the AllReduce. After the spectral convolution, the inverse partial DFT reconstructs each GPU's spatial output using the inverse basis $B_i^{-1} = \frac{1}{N_i} B_i^H$, applied in reverse dimension order. Because the AllGather has already produced the complete convolved spectrum on every GPU, this inverse requires no additional communication.

{\bf Spatial decomposition flexibility:}
DRIFT supports arbitrary $d$-dimensional spatial decompositions. A 1D slab decomposition partitions only one axis, which gives each GPU a local domain of $(N_x/P) \times N_y \times N_z$, and a 2D pencil decomposition partitions two axes, giving $(N_x/P_x) \times (N_y/P_y) \times N_z$ per GPU with $P = P_x P_y$. In both cases, the basis matrix (Eq.~\eqref{eq:basis_matrix}) adapts automatically through the phase offset $p_i N_i^{\mathrm{loc}}$, which shifts the DFT twiddle factors so that each GPU's local indices map to their correct global positions. Non-partitioned dimensions have $P_i = 1$ and zero offset. The partial spectrum and the subsequent AllReduce produce identical spectral coefficients, regardless of the decomposition topology. The choice of decomposition affects only the local tensor shape and the computational cost of the basis-matrix multiplication. The communication volume (Eq.~\eqref{eq:drift_volume}) is invariant to the decomposition, as it depends only on the mode count~$M$.

\begin{figure}[t]
    \centering
    \includegraphics[width=\columnwidth]{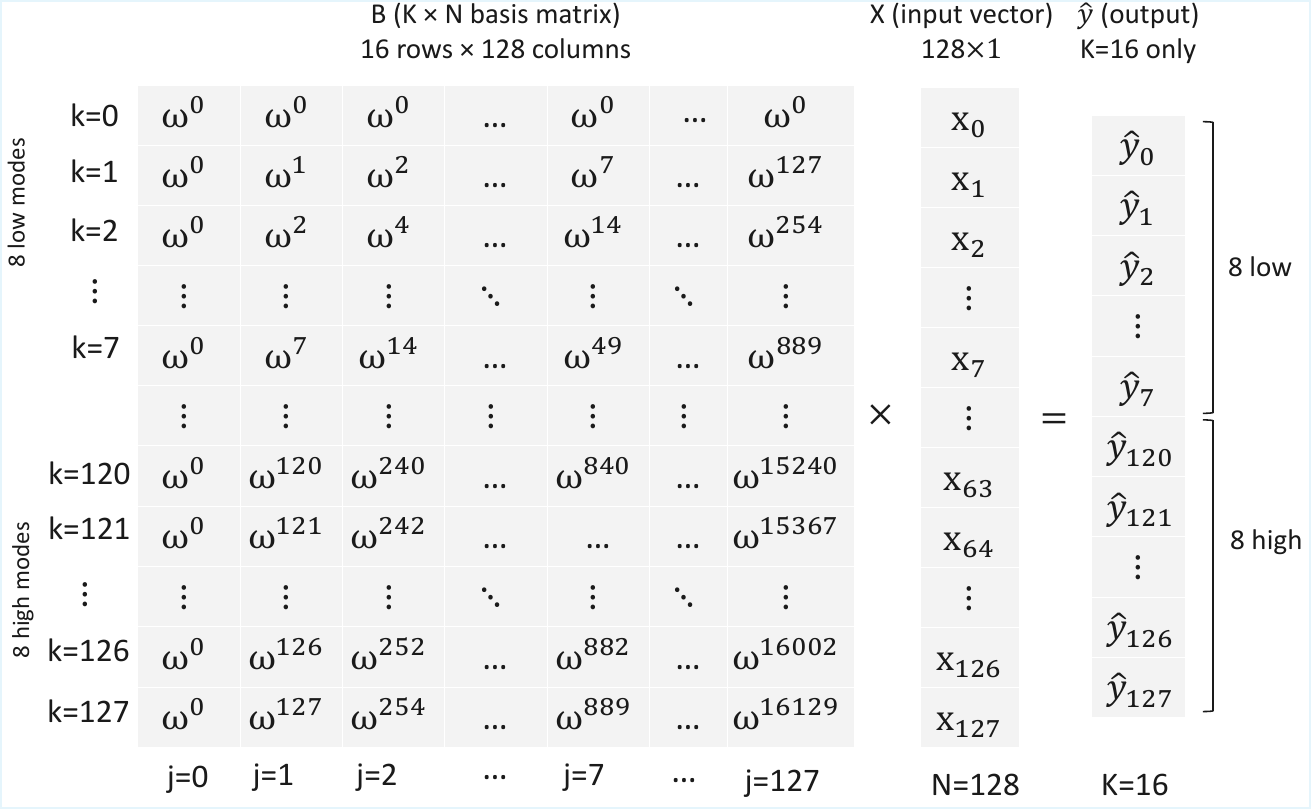}
    \caption{Partial DFT via \texttt{gemm} for one spatial dimension ($N{=}128$, $K{=}16$). The basis matrix $\mathbf{B} \in \mathbb{C}^{K \times N}$ with $B_{k,j} = \omega_N^{k \cdot j}$ contains only the rows corresponding to the $K$ frequency modes: 8~low ($k{=}0,\ldots,7$) and 8~high ($k{=}120,\ldots,127$).}
    \label{fig:basis-matrix}
\end{figure}

{\bf cuBLAS vs cuFFT:}
A natural question is whether replacing the FFT with dense matrix multiplication sacrifices computational efficiency. For a single 1-D transform of length~$N$ retaining $K = 2k_{\max}$ modes, the FFT requires $\frac{5}{2}N\log_2 N$ floating-point operations, but computes all $N$ frequencies, while the partial DFT via \texttt{gemm} requires $2KN$ operations for exactly the $K$ desired modes. Figure~\ref{fig:basis-matrix} illustrates the operation for one dimension, in which the basis matrix $\mathbf{B} \in \mathbb{C}^{K \times N}$ with entries $B_{k,j} = \omega_N^{k \cdot j}$ selects only the $K$ rows of the full DFT matrix and directly computes the desired spectral coefficients via a single \texttt{gemm} call. At $N = 128$ and $k_{\max} = 8$, the per-dimension \texttt{gemm} cost exceeds cuFFT's cost by a factor of~$1.8\times$.

\begin{figure}[t]
    \centering
    \includegraphics[width=\columnwidth]{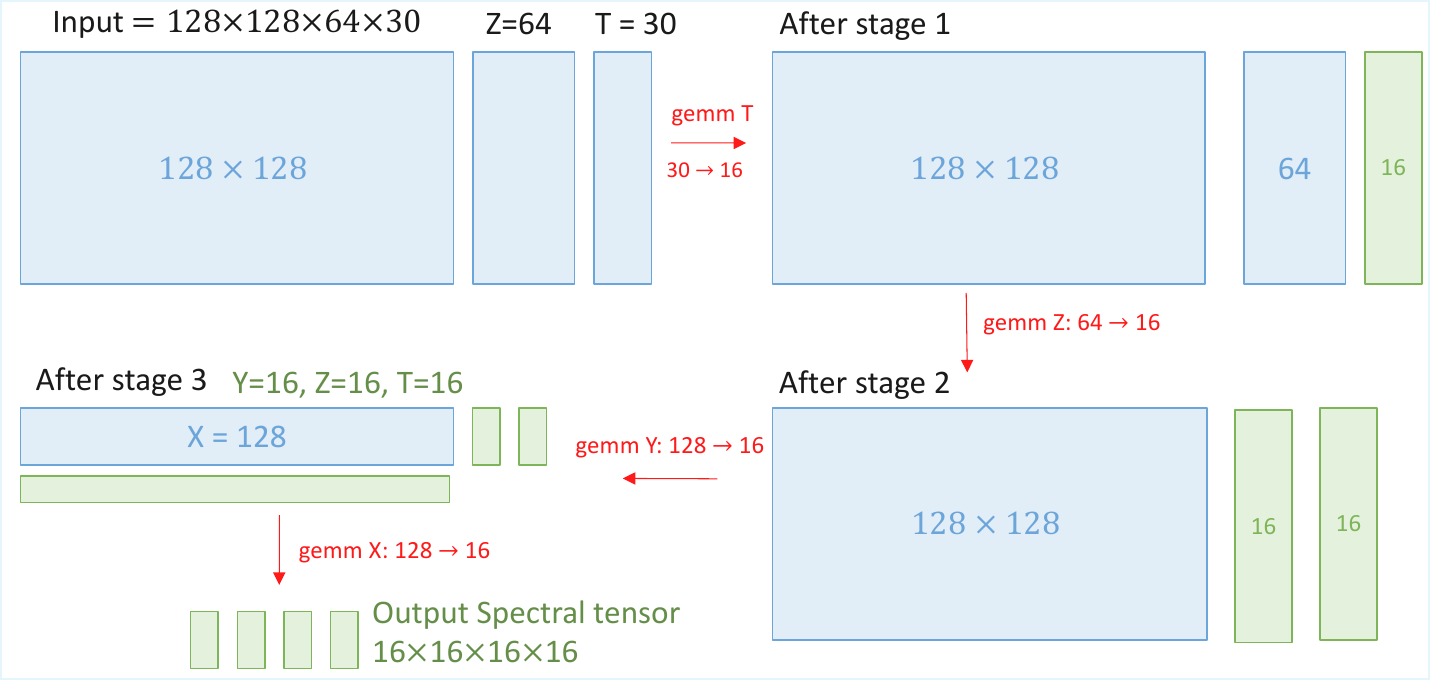}
    \caption{Progressive tensor compression across \texttt{gemm} stages. Each stage contracts one dimension (red arrow) via a basis-matrix multiply, reducing the tensor that subsequent stages operate on. Blue blocks denote full-sized dimensions; green blocks denote already-contracted dimensions ($K{=}16$). By stage~4, the tensor has been compressed in three dimensions, reducing the operational volume by $60\times$ relative to stage~1.}
    \label{fig:progressive-compression}
\end{figure}

However, when applied sequentially across all four dimensions, a \emph{progressive compression} effect makes the partial DFT cheaper than cuFFT overall (see Fig.~\ref{fig:progressive-compression}). Each stage contracts one dimension from $N_i$ to $K_i$, so the tensor feeding subsequent stages is progressively smaller. Stage~1 contracts $T$ from $30 \to 16$, operating on the full $128 \times 128 \times 64 \times 30$ tensor. After this contraction, stage~2 contracts $Z$ from $64 \to 16$ on a tensor already compressed along~$T$. Stage~3 contracts $Y$ from $128 \to 16$, with both $T$ and $Z$ already reduced, and stage~4 contracts $X$ from $128 \to 16$ on a tensor reduced in three dimensions, contributing only 1\% of the total \texttt{gemm} FLOPs (see Table~\ref{tab:flop-crossover}). For our grid ($128\times128\times64$, $T{=}30$, modes $(8,8,8,16)$), the total \texttt{gemm} computation across all four stages requires 13\% fewer floating-point operations than four equivalent cuFFT calls, and the advantage grows with resolution. Moreover, because the partial DFT maps each local axis of size $N_i^{\mathrm{loc}}$ directly to the $2k_{\max,i}$ retained modes, the full $N_i$-sized frequency representation is never allocated in GPU memory. This will eliminate the global memory traffic associated with writing and subsequently discarding the $(N_i - 2k_{\max,i})$ unused coefficients.

\begin{table}[t]
\centering
\caption{Per-stage \texttt{gemm} FLOPs with progressive compression vs.\ total cuFFT FLOPs. Grid $128{\times}128{\times}64$, $T{=}30$, modes $(8,8,8,16)$, width $d_v{=}20$.}
\label{tab:flop-crossover}
\renewcommand{\arraystretch}{1.2}
\begin{tabular}{llrrr}
\hline
Stage & Contracts & Tensor shape & \texttt{gemm} FLOPs & Share \\
\hline
1 (T) & $30 \to 16$  & $128{\times}128{\times}64{\times}30$  & 20.1\,G & 59\% \\
2 (Z) & $64 \to 16$  & $128{\times}128{\times}64{\times}16$  & 10.7\,G & 32\% \\
3 (Y) & $128 \to 16$ & $128{\times}128{\times}16{\times}16$  & 2.7\,G  & 8\%  \\
4 (X) & $128 \to 16$ & $128{\times}16{\times}16{\times}16$   & 0.3\,G  & 1\%  \\
\hline
\multicolumn{3}{l}{Total \texttt{gemm}} & 33.9\,G & \\
\multicolumn{3}{l}{Total cuFFT (no compression)} & 39.2\,G & \\
\multicolumn{3}{l}{\texttt{gemm}\,/\,cuFFT ratio} & \multicolumn{2}{l}{$0.87\times$ (13\% fewer)} \\
\hline
\end{tabular}
\end{table}

The dimension ordering in the progressive compression pipeline affects the total \texttt{gemm} cost. Contracting a large dimension early shrinks the tensor that all subsequent stages operate on. Table~\ref{tab:dim-ordering} compares the current implementation order (T, Z, Y, X) against the FLOP-optimal order (Y, Z, X, T) at $P{=}4$. Stage~1 costs the same in both cases because it always processes the full local tensor; the savings arise in stages~2--4, where the optimal order operates on an already-compressed tensor. Processing the largest local dimension first would reduce per-GPU \texttt{gemm} FLOPs by 30\%.

\begin{table}[t]
\centering
\caption{Dimension ordering: current vs.\ optimal per-GPU \texttt{gemm} FLOPs. Same configuration as Table~\ref{tab:flop-crossover}, $P{=}4$.}
\label{tab:dim-ordering}
\renewcommand{\arraystretch}{1.2}
\small
\begin{tabular}{c c r c r}
\hline
 & \multicolumn{2}{c}{Current} & \multicolumn{2}{c}{Optimal} \\
\cline{2-3}\cline{4-5}
Stage & Dim & GFLOPs & Dim & GFLOPs \\
\hline
1 & $T{:}30{\to}16$  & 20.1 & $Y{:}128{\to}16$ & 20.1 \\
2 & $Z{:}64{\to}16$  & 10.7 & $Z{:}64{\to}16$  &  2.5 \\
3 & $Y{:}128{\to}16$ &  2.7 & $X{:}32{\to}16$  &  0.6 \\
4 & $X{:}32{\to}16$  &  0.3 & $T{:}30{\to}16$  &  0.3 \\
\hline
 & Total & 33.9 & Total & 23.6 \\
\hline
\end{tabular}
\end{table}

\subsection{Backward Pass}
The forward pass uses only linear operations (basis-matrix multiplications) and two collectives (AllReduce and AllGather). During backpropagation, each collective reverses naturally. The backward pass of AllReduce is another AllReduce, and the backward pass of AllGather is an AllReduce followed by selecting the local partition. As a result, the backward pass performs the same collective operations as the forward pass and incurs identical communication cost. Because the spectral weights $R_\theta^{(p)}$ are partitioned, each GPU computes its weight gradients $\partial \mathcal{L} / \partial R_\theta^{(p)}$ locally with no additional gradient synchronization.

\subsection{Block Structure}
Each DRIFT block computes the full FNO residual $x^{(\ell+1)} = \sigma\bigl(W^{(\ell)} x^{(\ell)} + y^{(\ell)}\bigr)$, where $W^{(\ell)}$ is a point-wise linear operator applied independently on each GPU's local spatial partition and $y^{(\ell)}$ is the spectral convolution output reconstructed via the inverse partial DFT. Linear bypass requires only a single broadcast of the weight matrix from the root GPU, which contributes negligible communication relative to the spectral path. The complete per-block forward pass, including the dimension ordering that enables progressive compression (Table~\ref{tab:flop-crossover}), is summarized in Algorithm~\ref{alg:drift}.

\begin{algorithm}[t]
\caption{DRIFT block forward pass on GPU $p$}
\label{alg:drift}
\begin{algorithmic}[1]
\Require $x_p \in \mathbb{R}^{B \times d_v \times N_x^{\mathrm{loc}} \times N_y \times N_z \times N_t}$
\State $y_0 \gets W^{(\ell)} x_p$
\State $\hat{X}_p \gets \mathrm{pDFT}(x_p)$ along $T, Z, Y, X$
\State $\hat{X} \gets \texttt{AllReduce}(\hat{X}_p)$
\State $\hat{X}^{(p)} \gets \hat{X}[\mathcal{S}_p]$
\State $\hat{Y}^{(p)} \gets R_\theta^{(p)} \hat{X}^{(p)}$
\State $\hat{Y} \gets \texttt{AllGather}(\hat{Y}^{(p)})$
\State $y_p \gets \mathrm{piDFT}(\hat{Y})$ along $X, Y, Z, T$
\State \Return $\mathrm{GELU}(y_0 + y_p)$
\end{algorithmic}
\end{algorithm}

\section{Evaluation}
\subsection{Experimental Setup}

Table~\ref{tab:platform} summarizes the hardware and software platform used in this work. Each MPI rank is mapped to a single GPU, and all computation executes entirely on the GPU. Communication between GPUs uses GPU-aware MPI, which transfers data directly between GPU memory without staging through the host. All experiments use a 3D+time FNO, with a channel width $d_v = 20$, 4~Fourier blocks, and retained modes $(k_x, k_y, k_z, k_t) = (8, 8, 8, 16)$. Spatial data is distributed via a 1-D slab decomposition along the $x$-dimension. At $P{=}32$, where the local $x$-extent would fall below the mode count, a 2-D decomposition ($P_x{=}16$, $P_y{=}2$) is used instead. The $y$, $z$, and $t$ dimensions remain local on each GPU. Both DFNO~\cite{GradyDFNO} and DRIFT use identical model weights, inputs, and partitioning. The only difference is the spectral transform implementation. All timings reported are the mean over 20 iterations, following 5 warm-up steps.

\begin{table}[t]
\centering
\caption{Experimental platform.}
\label{tab:platform}
\small
\begin{tabular}{l l}
\toprule
\textbf{Parameter}          & \textbf{Value} \\
\midrule
GPU                         & NVIDIA Tesla V100-SXM2-32GB \\
GPUs per node               & 4 (NVLink) \\
Nodes                       & 1--8 (4--32 GPUs) \\
Inter-node network          & InfiniBand EDR (100 Gb/s) \\
\midrule
CUDA                        & 12.1 \\
Python                      & 3.10 \\
PyTorch                     & 2.1.0 \\
CuPy                        & 13.6.0 \\
mpi4py                      & 4.1.1 \\
MPI stack                   & HPC-X 2.19 (OpenMPI) \\
\bottomrule
\end{tabular}
\end{table}

\subsection{Dataset}
We evaluate on the 3D compressible Navier-Stokes dataset from PDEBench~\cite{PDEBench}, a widely adopted benchmark for scientific machine learning. The dataset contains 100 samples with random field initial conditions at Mach $M{=}1.0$ in the near-inviscid regime ($\eta{=}\zeta{=}10^{-8}$) on a periodic domain. Each trajectory has shape $(128, 128, 128, 21, 5)$ that is a $128^3$ spatial grid, 21~timesteps, and 5~field variables ($\rho, v_x, v_y, v_z, p$). We use a split of 90/10 for training/evaluation. The FNO is configured with $t_{\mathrm{in}}{=}5$ input timesteps and $t_{\mathrm{out}}{=}16$ predicted timesteps. For weak scaling experiments at smaller grid sizes, we slice the first $N_x$ points along the $x$-dimension from the native $128^3$ data.

\subsection{Numerical Exactness}
\begin{figure}[t]
  \centering
  \includegraphics[width=\columnwidth]{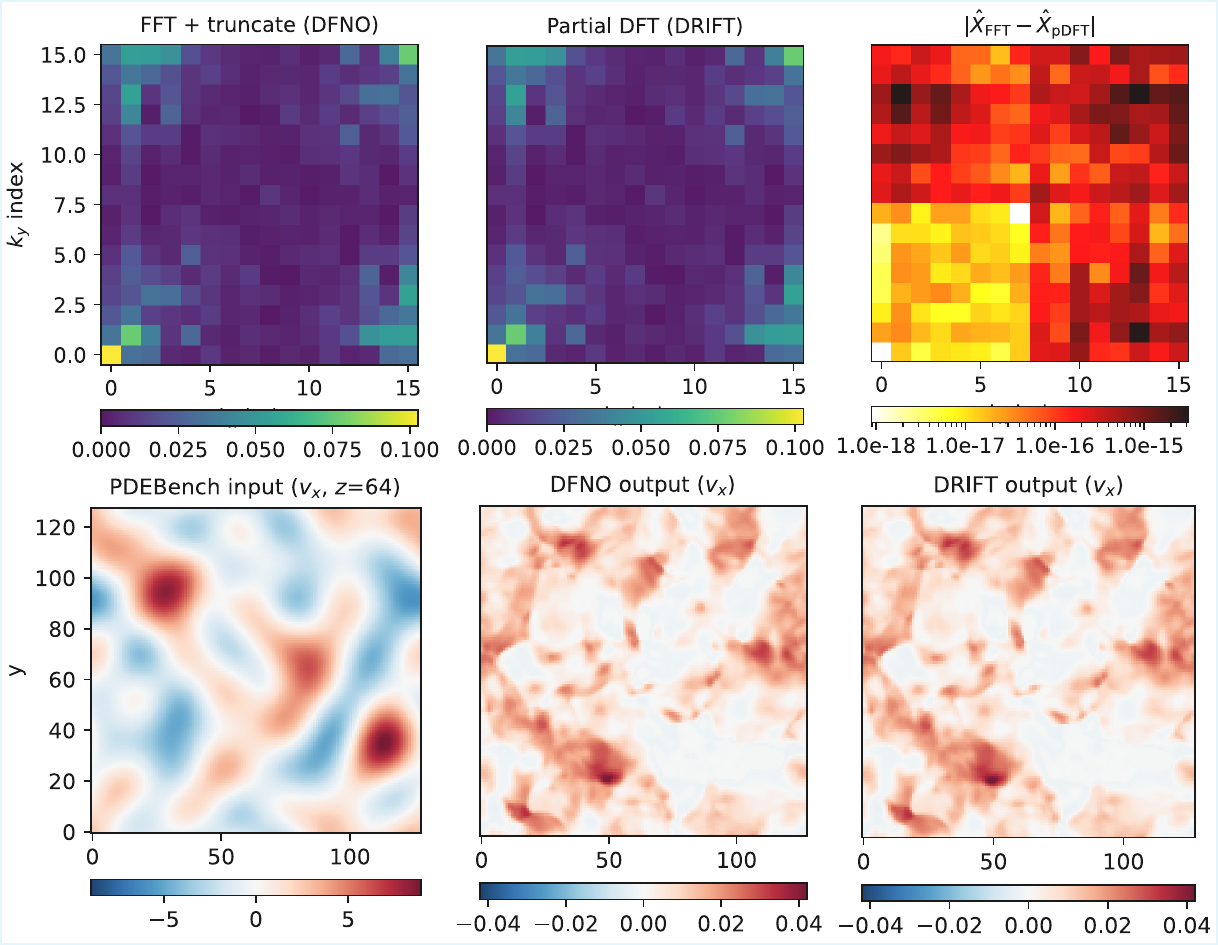}
  \caption{Above: Spectral coefficient comparison on PDEBench 3D compressible Navier-Stokes ($128^3$, $v_x$, $z{=}64$). Left to right: FFT + truncation (DFNO), partial DFT (DRIFT), pointwise absolute error. Relative Frobenius error:$3.2{\times}10^{-14}$. Below: Distributed full-model comparison ($P{=}16$ GPUs, matched weights). Input $v_x$ field, DFNO output, and DRIFT output at the $z{=}64$ midplane. The two models produce bit-wise identical results (relative $L_2 = 0$).}
  \label{fig:correctness}
\end{figure}

We verify that DRIFT reproduces the spectral coefficients of the standard FFT-based pipeline to machine precision. Figure~\ref{fig:correctness} compares the retained Fourier modes computed by FFT followed by truncation (DFNO) against the partial DFT basis matrix (DRIFT) on a PDEBench $v_x$ field at $128^3$ resolution. The two coefficient matrices are visually identical, and the relative Frobenius error is $3.2{\times}10^{-14}$, at the limit of double-precision arithmetic. Figure~\ref{fig:correctness} confirms that this exactness is preserved through the full distributed pipeline. Running DFNO and DRIFT with matched non-spectral weights on $P{=}16$ GPUs produces bitwise identical outputs (relative $L_2 = 0$).

\subsection{Strong Scaling}
\begin{figure}[t]
  \centering
  \includegraphics[width=\columnwidth]{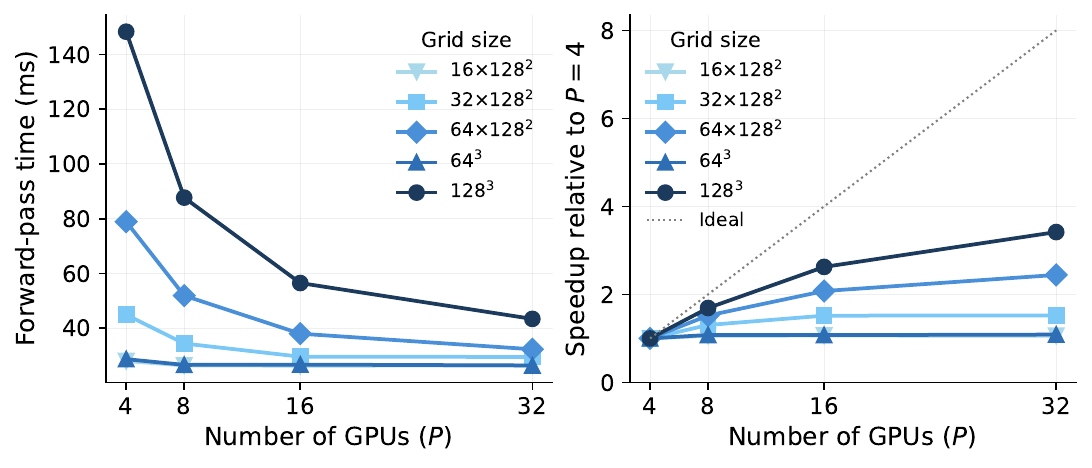}
  \caption{DRIFT strong scaling across five grid sizes. Left: forward-pass time. Right: speedup relative to $P{=}4$. Larger grids scale more efficiently as the increased local compute amortizes the fixed AllReduce cost. Modes $(8,8,8,16)$, $d_v{=}20$, 4 blocks, 4 GPUs per node, one rank per GPU.}
  \label{fig:strong_scaling}
\end{figure}

Table~\ref{tab:strong} reports forward-pass and backward-pass timings for DFNO and DRIFT on the PDEBench 3D compressible Navier-Stokes problem under strong scaling from 4 to 32~GPUs, for two different grid resolutions: $64^3$ and $128^3$. The $64^3$ grid is obtained by downsampling from the native $128^3$ data. On the $128^3$ grid, DRIFT achieves a 37.7--63.7$\times$ forward-pass speedup across all GPU counts. DFNO's forward time drops from 9443\,ms to 1634\,ms as $P$ increases from 4 to 32, a 5.8$\times$ reduction for an 8$\times$ increase in resources, as its all-to-all communication bottleneck limits scaling. DRIFT scales from 148.3\,ms to 43.4\,ms over the same range, a 3.4$\times$ reduction, with the absolute forward-pass time remaining under 150\,ms at all scales.
Figure~\ref{fig:strong_scaling} extends this analysis to five grid sizes derived from the native $128^3$ data. DRIFT's forward-pass time decreases with $P$ at all resolutions, and the $128^3$ grid tracks closest to ideal scaling, as its larger local volume provides more compute to amortize the fixed AllReduce cost. Smaller grids flatten earlier because the per-GPU volume shrinks and communication dominates sooner. The per-phase breakdown in Section~\ref{sec:phase_breakdown} quantifies this transition.
The backward pass exhibits a similar pattern. On the $128^3$ grid, DRIFT completes backpropagation in 49.8\,ms at $P{=}32$ compared to DFNO's 1691.5\,ms, which yields a 35.7$\times$ total (forward + backward) speedup relevant for training workloads.

\begin{table}[t]
\centering
\caption{DRIFT vs.\ DFNO strong scaling on PDEBench 3D compressible Navier-Stokes. Modes $(8,8,8,16)$, width~20, 4~blocks, 4 GPUs per node, one rank per GPU.}
\label{tab:strong}
\small
\setlength{\tabcolsep}{4pt}
\begin{tabular}{clrrrrr}
\toprule
 & & \multicolumn{2}{c}{Fwd (ms)} & \multicolumn{2}{c}{Bwd (ms)} & \\
\cmidrule(lr){3-4} \cmidrule(lr){5-6}
Grid & $P$ & DFNO & DRIFT & DFNO & DRIFT & Speedup (fwd) \\
\midrule
\multirow{4}{*}{$64^3$}
 &  4 &  926.6 & 28.7 &  954.9 & 38.9 & 32.3$\times$ \\
 &  8 &  488.6 & 26.6 &  483.2 & 28.9 & 18.4$\times$ \\
 & 16 &  300.5 & 26.7 &  249.1 & 24.1 & 11.3$\times$ \\
 & 32 &  226.2 & 26.4 &  159.7 & 23.5 &  8.6$\times$ \\
\midrule
\multirow{4}{*}{$128^3$}
 &  4 & 9443.1 & 148.3 & 10140.7 & 246.8 & 63.7$\times$ \\
 &  8 & 5154.8 &  87.7 &  5505.1 & 133.1 & 58.8$\times$ \\
 & 16 & 3202.6 &  56.5 &  3259.2 &  77.2 & 56.7$\times$ \\
 & 32 & 1634.1 &  43.4 &  1691.5 &  49.8 & 37.7$\times$ \\
\bottomrule
\end{tabular}
\end{table}

\subsection{Per-Phase Breakdown}
\label{sec:phase_breakdown}
\begin{figure}[t]
  \centering
  \includegraphics[width=0.9\columnwidth]{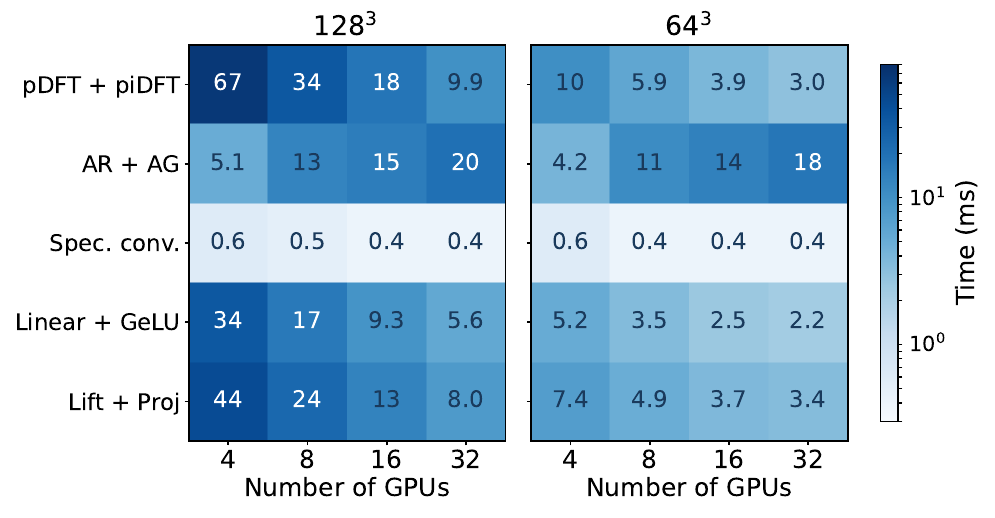}
  \caption{Per-block phase breakdown for DRIFT at $128^3$ (left) and $64^3$ (right) grids. Modes $(8,8,8,16)$, $d_v{=}20$, 4 GPUs per node, one rank per GPU.}
  \label{fig:phases}
\end{figure}

Figure~\ref{fig:phases} shows the per-block timing breakdown for DRIFT on $128^3$ and $64^3$ grids across GPU counts. Each heatmap decomposes a single spectral convolution block into five phases: forward and inverse partial DFT, AllReduce plus AllGather communication, spectral convolution, skip-connection linear projection with GeLU activation, and the lift/projection layers.

On the $128^3$ grid at $P{=}4$, the forward and inverse partial DFT dominate at 67\,ms (44\% of total per-block time), reflecting the large local spatial volume ($32{\times}128{\times}128$ per GPU). Communication accounts for only 5.1\,ms (3.4\%). As $P$ increases to 32, the local volume shrinks and the partial DFT cost drops to 9.9\,ms, while AllReduce plus AllGather grows to 20\,ms and becomes the dominant phase. This transition from compute-bound to communication-bound explains the decreasing speedup over DFNO at higher GPU counts in Table~\ref{tab:strong}. Even so, the absolute communication time remains below 20\,ms, over 44$\times$ lower than DFNO's 891\,ms.

The spectral convolution remains negligible at all scales (0.4--0.6\,ms) since it operates on the small $K_x{\times}K_y{\times}K_z{\times}K_t$ coefficient tensor, regardless of the spatial grid size. The $64^3$ grid shows a similar transition, though shifted to lower $P$. Communication already dominates at $P{=}8$ (11\,ms out of total 26\,ms), consistent with the smaller per-GPU volume, providing less compute to amortize the collective cost. Notably, DRIFT's communication time is nearly identical across the two grid sizes (5.1\,ms vs.\ 4.2\,ms at $P{=}4$; 20\,ms vs.\ 18\,ms at $P{=}32$), confirming that the AllReduce payload depends only on the retained mode count, not on the spatial resolution (Eq.~\eqref{eq:drift_volume}).

\subsection{Weak Scaling}
\begin{figure}[t]
\centering
\includegraphics[width=\columnwidth]{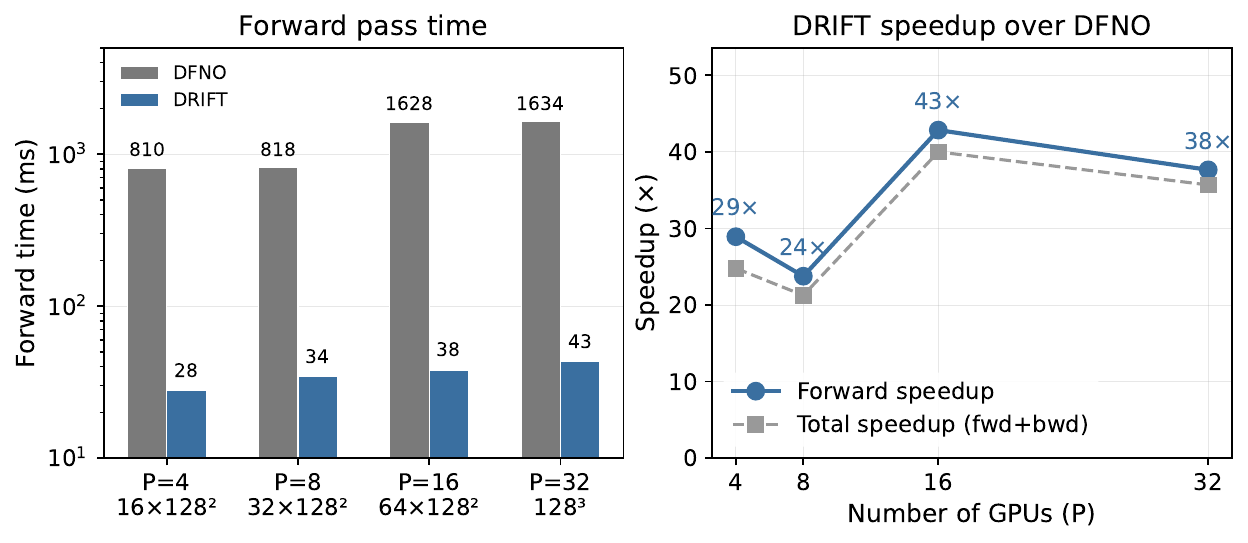}
\caption{Weak scaling of DRIFT vs. DFNO. Left: forward-pass time; right: speedup. The local tensor size per GPU is fixed at $4{\times}128{\times}128$, and the global grid grows with $P$ along the $x$-dimension from $16{\times}128{\times}128$ ($P{=}4$) to $128^3$ ($P{=}32$). Modes $(8,8,8,16)$, $d_v{=}20$, 4 blocks, 4 GPUs per node, one rank per GPU.}
\label{fig:weak}
\end{figure}

Figure~\ref{fig:weak} presents weak scaling, where the local tensor size per GPU is held constant at $4{\times}128{\times}128$, while the global grid grows with $P$ along the distributed $x$-dimension, from $16{\times}128{\times}128$ at $P{=}4$ to $128^3$ at $P{=}32$. DRIFT's forward-pass time increases from 28\,ms to 43\,ms as $P$ grows from 4 to 32, a 1.5$\times$ increase for an 8$\times$ growth in problem size. DFNO, by contrast, grows from 810\,ms to 1634\,ms over the same range, as each increase in $P$ introduces additional all-to-all communications. The resulting forward speedup peaks at 43$\times$ at $P{=}16$ and remains above 24$\times$ across all configurations. The speedup rises from 29$\times$ at $P{=}4$ to 43$\times$ at $P{=}16$, then drops to 38$\times$ at $P{=}32$. The initial increase reflects DFNO's communication cost scaling super-linearly with $P$ in the 1D decomposition regime, amplifying the gap. The drop at $P{=}32$ coincides with the transition to a 2D decomposition ($P_x{=}16$, $P_y{=}2$), which reduces DFNO's per-collective message size and partially alleviates its communication bottleneck, while DRIFT's AllReduce cost continues to grow with $P$.
DRIFT's forward time is dominated by local partial-DFT and spectral convolution compute at low $P$, with AllReduce communication becoming the dominant cost at the largest GPU counts. This trend contrasts with DFNO's behavior, where communication accounts for 95--98\% of forward-pass time at every scale. The total speedup (forward plus backward) tracks the forward speedup closely, which ranges from 24$\times$ to 38$\times$, and indicates that DRIFT's advantages extend to training workloads.

\subsection{Sensitivity to $k_{\max}$}
\begin{figure}[t]
\centering
\includegraphics[width=\columnwidth]{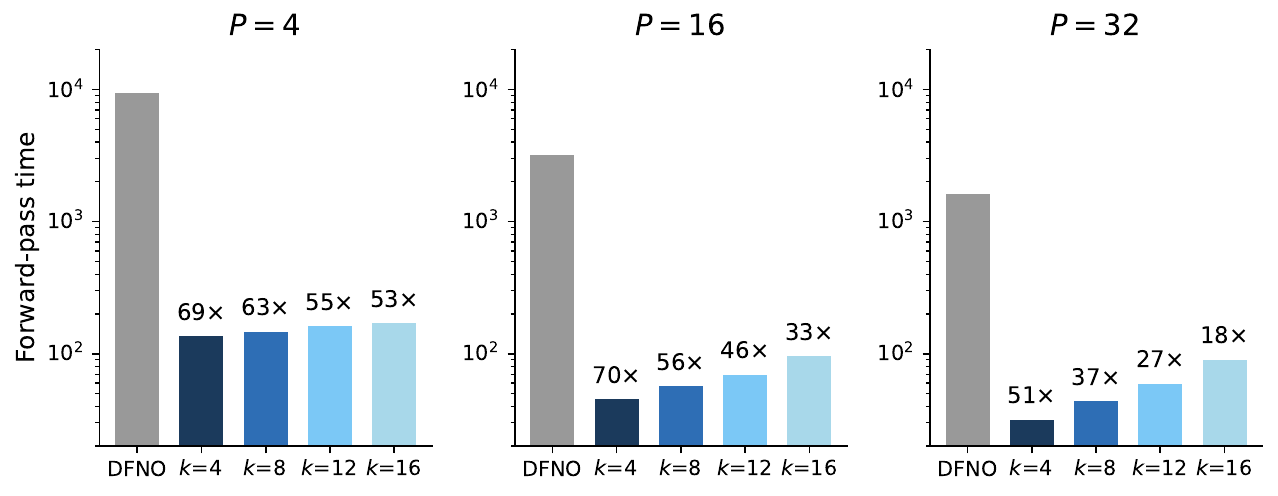}
\caption{Effect of $k_{\max}$ on forward-pass time and speedup across 4, 16, 32 GPU counts. DRIFT's advantage grows as fewer modes are retained. $128^3$ grid, $d_v{=}20$, 4 blocks, 4 GPUs per node, one rank per GPU.}
\label{fig:kmax}
\end{figure}

Figure~\ref{fig:kmax} shows the sensitivity of DRIFT's forward-pass time and speedup to the number of retained modes on the PDEBench $128^3$ grid across three GPU counts. As $k_{\max}$ increases, DRIFT's communication payload (Eq.~\eqref{eq:drift_volume}) and partial DFT cost both grow, which reduces the speedup over DFNO. At $P{=}4$, the speedup ranges from 69$\times$ at $k{=}4$ to 53$\times$ at $k{=}16$, a modest decline because the local partial DFT dominates at low $P$ and scales linearly with $k_{\max}$. At $P{=}32$, the effect is more pronounced. The speedup drops from 51$\times$ at $k{=}4$ to 18$\times$ at $k{=}16$, as the AllReduce payload grows with $k_{\max}^d$ and communication already dominates at this GPU count. Crucially, DFNO's forward-pass time is independent of $k_{\max}$ because it communicates and computes the full spatial tensor regardless of mode count. DRIFT maintains a substantial speedup across all configurations, confirming that the advantages hold when a large fraction of modes is retained.

\subsection{Training Convergence}
\begin{figure}[t]
  \centering
  \includegraphics[width=\columnwidth]{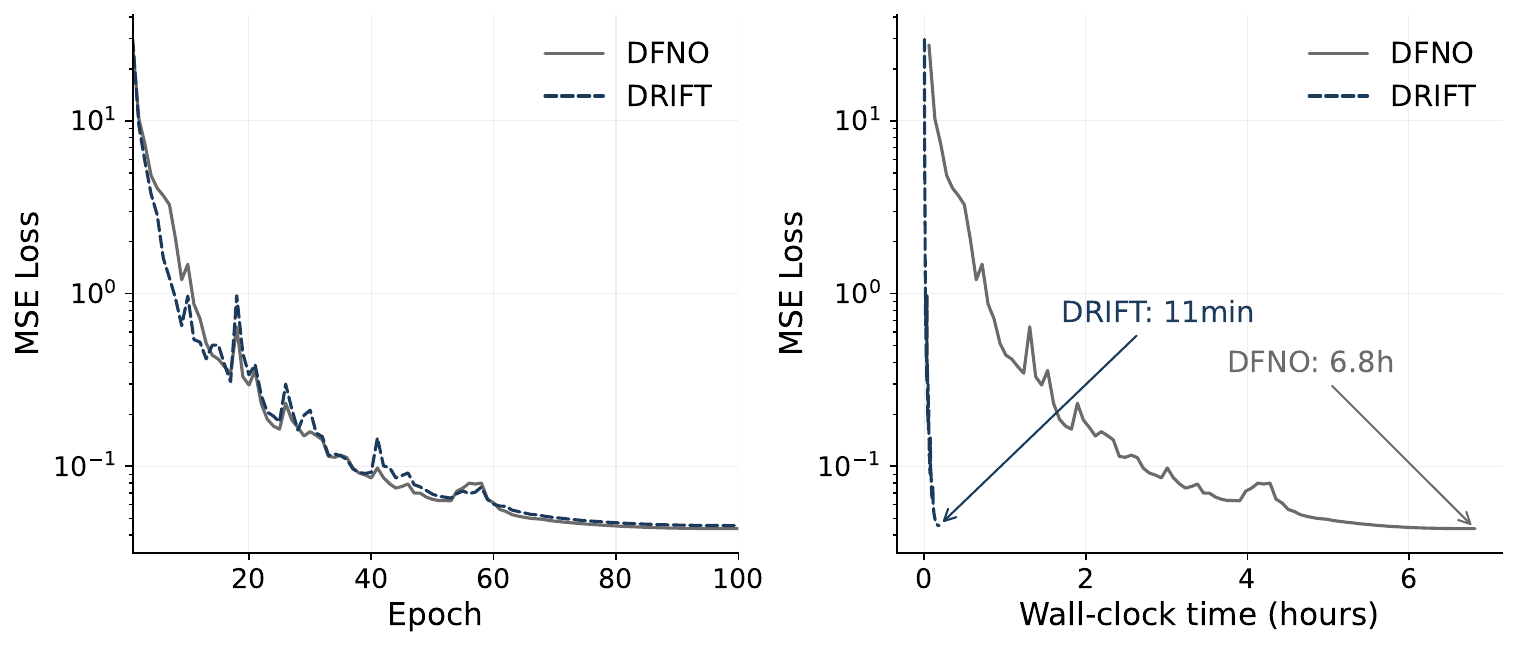}
  \caption{Training convergence on PDEBench 3D compressible Navier-Stokes ($128^3$, $P{=}16$ GPUs, 4 GPUs per node, one rank per GPU). Left: training loss vs.\ epoch. Right: training loss vs.\ wall-clock time.}
  \label{fig:training}
\end{figure}

The preceding experiments measured individual forward and backward passes in isolation. However, end-to-end training includes additional overhead from optimizer steps, gradient accumulation, data loading, and memory management. To verify that DRIFT's advantages hold in practice, we trained both DFNO and DRIFT on the PDEBench $128^3$ compressible Navier-Stokes dataset for 100 epochs run on $P{=}16$ GPUs, using 40 of the 90 training samples and all 10 test samples. Both methods minimize MSE loss using the Adam optimizer, with a learning rate of $10^{-3}$ and batch size of~1. Both models use identical architectures (4 blocks, $d_v{=}20$, modes $(8,8,8,16)$), identical random initializations, and the same sample ordering per epoch. After each epoch, we evaluate the relative $L_2$ error $\|y - \hat{y}\|_2 / \|\hat{y}\|_2$ on the test set.
Figure~\ref{fig:training} shows training loss versus epoch (left) and versus wall-clock time (right). Both methods converge to comparable final training loss (DFNO: 0.044, DRIFT: 0.045) and relative $L_2$ error (DFNO: 0.23, DRIFT: 0.25). The similar trends in training loss and $L_2$ error confirm that DRIFT is a drop-in replacement that does not affect model quality. Both methods achieve a training loss below 0.1 at epoch~36, but DRIFT arrives there in 4~minutes, while DFNO requires 2.6~hours. DRIFT averages 6.6\,s per epoch, compared to DFNO's 245\,s, and completes 100 epochs in 11 minutes versus 6.8 hours, a $37\times$ wall-clock speedup.

\section{Discussion}
{\bf Scaling projections:} The communication models (Eqs.~\ref{eq:cost_drift} and~\ref{eq:cost_dfft}) predict how each approach scales. The all-to-all cost in Eq.~\ref{eq:cost_dfft} has a bandwidth term that decreases at a rate of $1/P$, which is why DFNO's communication time drops with $P$ in the strong scaling (see Table~\ref{tab:strong}) and the per-phase breakdown (Section~\ref{sec:phase_breakdown}). At our scale, all-to-all remains bandwidth bound. However, the latency term grows as $O(P)$ and is predicted to dominate at larger GPU counts. By contrast, the AllReduce cost in Eq.~\ref{eq:cost_drift} has a fixed bandwidth term proportional to $M$ and a latency term that grows only as $O(\log P)$, consistent with the modest increase in DRIFT's communication time observed. At larger scales, the all-to-all latency grows as $O(P)$, while the AllReduce latency grows as $O(\log P)$, suggesting that DRIFT's communication advantage will continue to increase with GPU count.

{\bf Applicability beyond FNO:} DTST can be applied to any spectral method that retains a subset of frequency modes from distributed data, including spectral element methods~\cite{KomatitschVilotte1998}, spherical harmonic transforms in climate models~\cite{Wedi2013}, and neural architectures with truncated spectral filters. The only requirement is that $\mathcal{S}$ is known at construction time.

{\bf Crossover regime:}
DRIFT's computational advantages rely on $k_{\max} \ll N$. The \texttt{gemm} path becomes more expensive than cuFFT at $k_{\max} \approx 10$ ($k_{\max}/N \approx 0.08$), but typical FNO configurations use $k_{\max} = 4$--$12$, where DRIFT's partial DFT remains more efficient than the full transform. Even beyond this crossover, DRIFT's communication advantage persists.

{\bf Limitations:}
DRIFT's AllReduce payload is fixed at $O(M)$, regardless of value of $P$, so the communication fraction increases as the per-rank compute shrinks. At sufficiently large $P$, DRIFT becomes communication-bound and further scaling yields diminishing returns. Scaling behavior at hundreds of GPUs may also be affected by network topology and congestion, which is not captured by the $\alpha$--$\beta$ model. Additionally, we evaluate on one FNO architecture, and measured speedups may differ for other problems or larger channel widths.

\section{Conclusion}
In this work, we have introduced the Distributed Truncated Spectral Transform (DTST), a communication primitive that replaces the distributed FFT with local partial DFTs and two collectives on retained spectral coefficients. DTST reduces per-layer communication volume from $O(N^d / P)$ to $O(M)$, with $O(\log P)$ latency scaling, while producing identical spectral coefficients. We presented DRIFT, a GPU implementation for distributed Fourier Neural Operators using separable per-dimension partial DFTs via cuBLAS and GPU-aware MPI. DRIFT avoids both the unnecessary communication and the unnecessary computation inherent in the standard distributed FFT by computing only the needed frequency modes and communicating only the retained spectrum. On a 3D+time FNO across 4--32 GPUs, DRIFT achieves a 38--64$\times$ forward-pass speedup and a 37$\times$ training speedup over the distributed FNO baseline, with comparable convergence. DRIFT's advantages grow with both spatial resolution and problem dimensionality.


\balance

\end{document}